\documentclass[10pt, conference, letterpaper]{IEEEtran}
\IEEEoverridecommandlockouts
\usepackage{cite}
\usepackage{amsmath,amssymb,amsfonts}
\usepackage{algorithm}
\usepackage{algpseudocode}
\usepackage{array}
\usepackage[caption=false,font=normalsize,labelfont=sf,textfont=sf]{subfig}
\usepackage{textcomp}
\usepackage{stfloats}
\usepackage{url}
\usepackage{verbatim}
\usepackage{graphicx}
\usepackage{hyperref}
\usepackage{xcolor}
\usepackage{soul}
\graphicspath{ {./} }

\def\BibTeX{{\rm B\kern-.05em{\sc i\kern-.025em b}\kern-.08em
		T\kern-.1667em\lower.7ex\hbox{E}\kern-.125emX}}

\def\BibTeX{{\rm B\kern-.05em{\sc i\kern-.025em b}\kern-.08em
		T\kern-.1667em\lower.7ex\hbox{E}\kern-.125emX}}

\begin{document}
	
	\title{Attack Mitigation in Gateways of Pervasive Systems}
	
\author{Erol Gelenbe\thanks{Erol Gelenbe is with the Institute of Theoretical \& Applied Informatics, Polish Academy of Science, 44-100 Gliwice, PL, 
Universit\'{e} C\^{o}te d'Azur, I3S CNRS, 06100 Nice, FR, and the Department of Engineering at King's College London, WC2R 2LS, UK.
ORCID: 0000-0001-9688-2201, email: gelenbe.erol@gmail.com},~\IEEEmembership{Fellow,~IEEE,} and Mohammed Nasereddin\thanks{Mohammed Nasereddin is with the Institute of Theoretical \& Applied Informatics, Polish Academy of Science, 44-100 Gliwice, PL,
ORCID: 0000-0002-3740-9518, email: mnasereddin@iitis.pl}}

\maketitle

\begin{abstract}
 In pervasive systems, mobile devices and other sensors access Gateways,  which are Servers that communicate with the devices,  provide low latency services, connect them with each other, and connect them to the Internet and backbone networks. Gateway Servers are often equipped with Attack Detection (AD) software that analyzes the incoming traffic to protect the system against Cyberattacks, which can overwhelm the Gateway and the system as a whole. This paper describes a traffiic shaping, attack detection and an optimum attack mitigation scheme to protect the Gateway and the system as a whole from Cyberattacks. The approach is described and evaluated in an experimental test-bed. The key parameter of the optimum mitigation technique is chosen based on an analytical model whose predictions are validated through detailed experiments.
\end{abstract}

\section{Introduction}
Pervasive systems that use some of the Internet's $30$ Billion devices and Internet users, are subject to Cyberattacks \cite{Cisco2020,liu2017novel} which compromise the victim nodes and disable systems through malware and packet floods \cite{Cyber23}. For instance, in $2017$, an attack  took down $180,000$ servers with $2.54$ Tera-bits/sec of traffic \cite{Cloudflare1}. Such attacks \cite{Spilios} compromise  their victims and  also turn them into attackers \cite{Sinanovic}, creating packet Floods that cause congestion, system crashes \cite{Iqbal}, and even overload the Attack Detection (AD) systems. Therefore it is crucial to protect our pervasive systems with effective AD and mitigation techniques. 

Thus,  in Section \ref{Lit} we briefly review the literature on leading-edge AD and Mitigation. Then, in Section \ref{sec:Methodology}, we introduce a novel approach that combines traffic shaping to protect the AD from attack overloads, together with periodic AD that examines  the incoming traffic, and adaptively drops packets when an attack is detected. This approach is based on an Adaptive Attack Mitigation (AAM) algorithm that minimizes a cost function combining the overhead of the AD when it examines packets, and the loss of a fraction of the benign packets due to the packet drops when an attack is detected. This approach is then evaluated with extensive measurements, showing a good agreement between the theoretical results 
in Section \ref{sec:Methodology} and the experimental measurements in Section \ref{sec:Results}. 
Conclusions and ideas for future work are presented in Section \ref{sec:Conclusion}.

\subsection{Literature Review} \label{Lit}
Attack detection and mitigation systems often employ machine learning (ML) and deep learning (DL) techniques, such as Decision Trees \cite{he2020machine}, Convolutional Neural Networks (CNNs) \cite{aldhyani2022attacks}, and Generative Adversarial Networks (GANs) \cite{kavousi2020evolutionary}, to identify anomalies and detect cyber threats. Most evaluations are conducted off-line and do not consider the real-time impact of attacks  and the consequence of the attack on a given system \cite{Banerjee,Tuan,Al-Issa,Guven,gelenbe2022traffic,nakip2024associated}, however a few test-beds have been 
constructed for experimentation with attacks to observe their effect on a given system \cite{Kaouk2018,annor2018development,waraga2020design}.

Several security protocols and architectures have been developed to meet the specific demands of Gateway environments, with emphasis on securing critical autonomous vehicles \cite{kim2021cybersecurity}, and the 6G-enabled Internet of Vehicles (IoV) \cite{sedjelmaci2023secure}. Recent advances also emphasize collaborative and adaptive solutions to address data-sharing limitations across heterogeneous Gateways. For instance, transfer \cite{li2021transfer} and federated learning \cite{DISFIDA} enable knowledge transfer between devices and networks without compromising data privacy, and achieve significant improvements in attack detection rates. Advanced AD systems can incorporate innovative techniques such as spatial-temporal analysis, Bayesian networks \cite{aloqaily2019intrusion}, and hybrid approaches to secure smart cities \cite{aurangzeb2023cybersecurity, pascale2021cybersecurity} and critical SCADA systems \cite{ghaleb2016scada,tesfahun2016scada}. Furthermore, integrating DL models with edge computing and Multi-access Edge Computing (MEC) has been useful in real-time attack mitigation \cite{alladi2021artificial}. Ensemble learning and proactive detection methods using Bayesian DL and Discrete Wavelet Transform, have shown the importance of adaptive techniques to counter threats targeting Gateways \cite{eziama2020detection}.

\section{Methodology}\label{sec:Methodology}
\begin{figure}[h!]
	\centering
	\includegraphics[height=4cm,width=9.5cm]{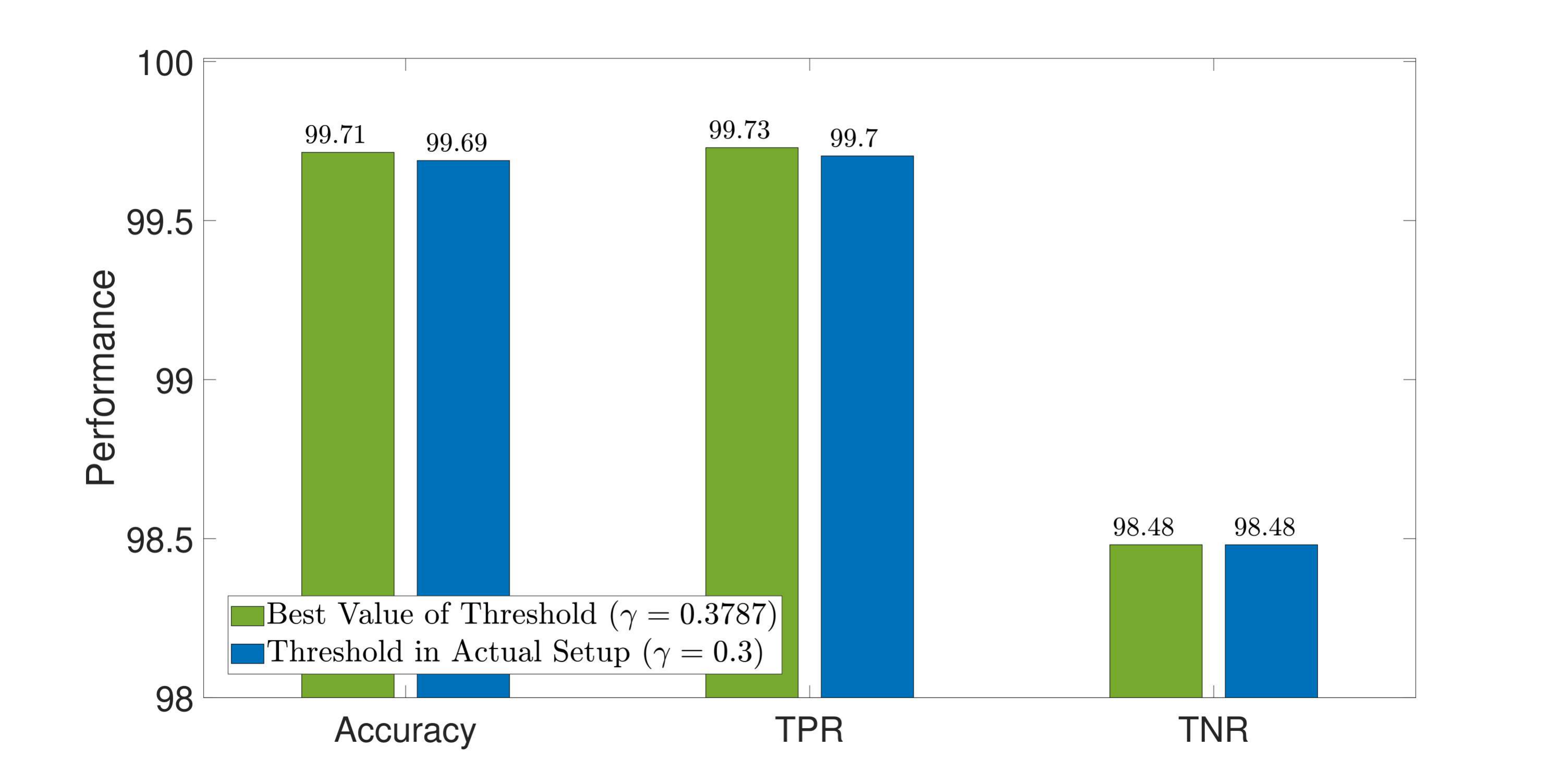}
	\caption{The performance of the AADRNN attack detector that was evaluated on the test-bed \cite{LCN23}.}
	\label{Accuracy}
\end{figure}

In this section, we present the proposed Adaptive Attack Mitigation (AAM) system. It consists 
of AD, followed by  a Smart QDTP Forwarder (SQF) which protects the Server from excessive packet backlogs during a Flood Attack, and the novel AAM algorithm which is  triggered when the attack detector identifies an attack and rapidly drops malicious packets. This algorithm minimizes a cost function $C(AAM)$ which combines the overhead of testing incoming packet streams and dropping benign as well as malicious packets.
The system includes an AD system installed in the Gateway Server  \cite{Brun} using the Random Neural Network \cite{RNN} that was validated in \cite{gelenbe2023iot}. The AD performance was evaluated both with the Kitsune attack dataset \cite{kitsune_paper,kitsune_dataset}, and on an experimental test-bed \cite{LCN23}, exhibiting in Figure \ref{Accuracy} its high accuracy of $99.69\%$ with a True Positive Rate (TPR) of $99.71\%$, and a True Negative Rate (TNR) of $98.48\%$. 

However, relying solely on the AD is insufficient for system protection. Indeed, Flood attacks lead to a huge packet accumulation at the AD, causing large queueing delays as illustrated by the red curve in Figure \ref{QL-10s&60s} (above). For instance, a 10-second attack floods the Server with over $153,667$ malicious packets mixed with normal traffic from various devices in the network and overwhelms the AD processing capacity. During longer 60-sec attacks, Figure \ref{QL-10s&60s} (below) shows that the attack causes over $400,000$ packets to accumulate and severely congest the AD, leading to the paralysis of the resource-constrained Gateway Server and prolonged downtimes or even system failure.

\subsection{The Smart QDTP Policy Forwarder (SQF)}

\begin{figure}[t!]
	\centering
	\includegraphics[height=5cm,width=8.5cm]{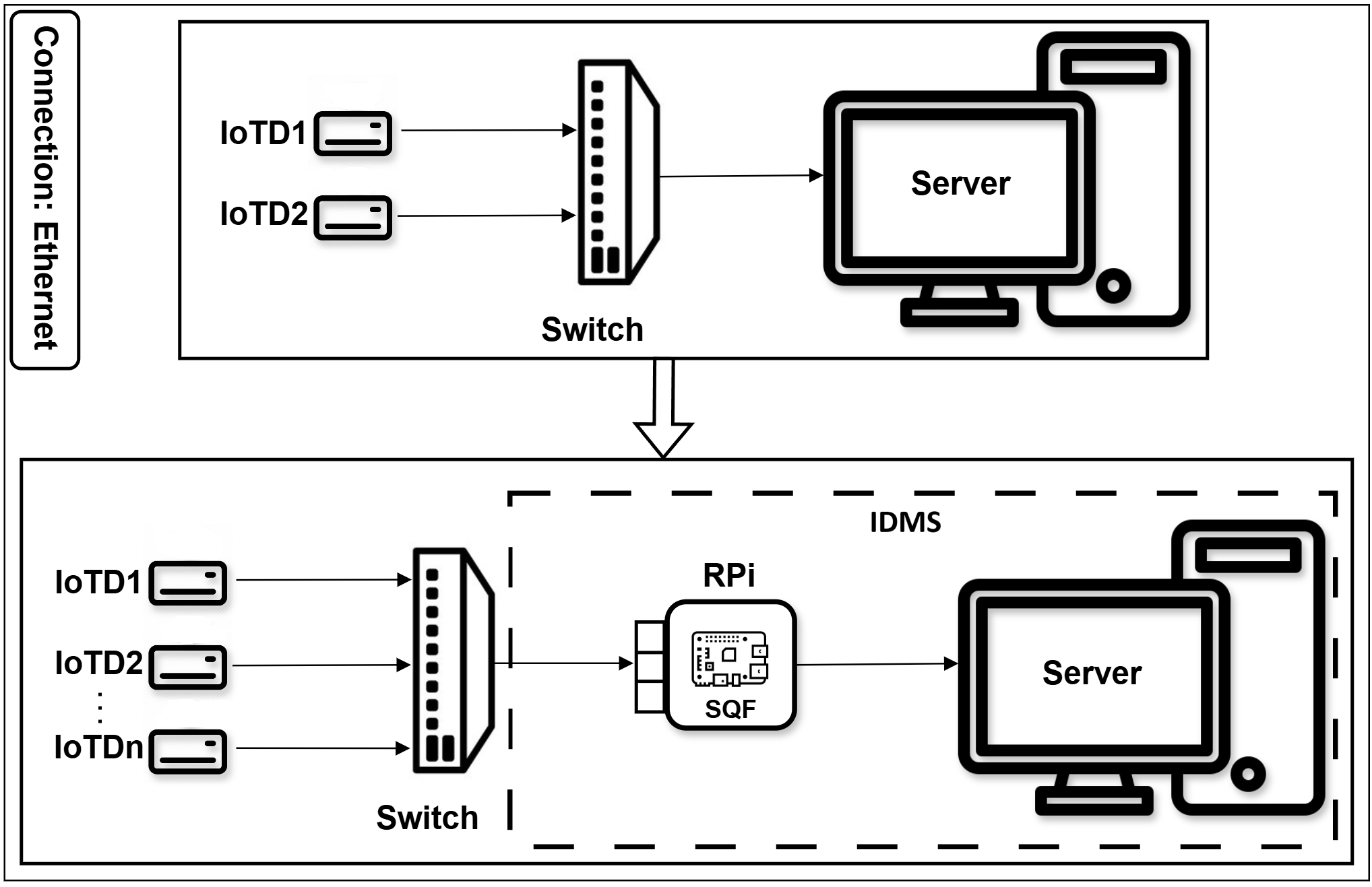}\\
	\caption{The experimental test bed consists of Gateway devices connected directly to the Server via a switch using Ethernet (above). In the modified architecture, the SQF is placed between the Server and the Gateway devices, isolating the Server and acting as a traffic-shaping interface (below).}
	\label{Zero-0}
\end{figure}

Thus, to protect the Gateway Server from being paralysed by the congestion that occurs when it is targeted by a Flood Attack,  the initial system architecture shown in Figure \ref{Zero-0} (above) was modified 
to incorporate traffic shaping with the quasi-deterministic transmission policy (QDTP)  \cite{ICC22} that we install on a low-cost Raspberry Pi.  The resulting architecture with the smart QDTP forwarder (SQF), is shown in Figure \ref{Zero-0} (below). SQF forwards the $n$-th arriving packet that arrives at time $a_n,~ n\geq 0$, to the Server at time $t_n$ defined by $t_0 = a_0$ and:
\begin{eqnarray}
&&t_{n+1}=\max(t_n+D,a_{n+1}), ~n\geq 0,\label{eq1}\\
&&hence:~t_{n+1}-t_n\geq D~, \label{eq2}
\end{eqnarray}
where $D > 0$ is a constant parameter. Thus, the total delay $Q_n$ experienced by the $n$-th packet  is given by:
\begin{eqnarray}
&&Q_0=t_0-a_0=0,~Q_{n+1}=t_{n+1}-a_{n+1},\nonumber\\
&&=\max(t_n+D,a_{n+1})-a_{n+1},\nonumber\\
&&=0,~if~t_n+D\leq a_{n+1},~and\nonumber\\
&&=t_n+D-a_{n+1},~otherwise.
\end{eqnarray}
Since $t_n=Q_n+a_n$, we obtain the recursive expression:
\begin{eqnarray}
&&Q_{n+1}=\max(0,t_n+D-a_{n+1}),\nonumber\\
&&=\max(0,Q_n+D-a_{n+1}),~n\geq 0.\label{Q}
\end{eqnarray}
We have measured the actual transmission time from the Raspberry Pi to the Gateway Server, including 
the Simple Network Management Protocol (SNMP) time taken at the Server, and it is less than $15\%$  of the approximately $\tau\approx 3$ milliseconds taken by the AD to process one packet.  
Therefore, when the $n$-th packet is forwarded by the SQF, we can assume that it instantly reaches the Server’s input queue for AD processing.  The effect of the SQF approach was  experimentally validated in \cite{BestPaper}, and the blue curve in Figures \ref{QL-10s&60s} (above) and (below), clearly show that the SQF successfully protects the Server during a Flood Attack, resulting in a Server queue length that remains very short, both during and after the attack. 

\subsection{Adaptive Attack Mitigation (AAM)}\label{subsec:AAM}
The blue curve in Figures \ref{QL-10s&60s} (above) and (below), clearly shows that the SQF successfully protects the Server from severe packet accumulation during a Flood Attack, maintaining the queue length of the Server at normal levels both during and after the attack. However, while this approach safeguards the Server from being overwhelmed, it does not stop attack packet flow; instead, these packets accumulate at the entrance of the SQF, creating a substantial backlog that must eventually be processed by the Server. Thus, although the Server is protected from paralysis, challenges such as high delays and benign packet loss persist.

To address these challenges, we introduce a novel Adaptive Attack Mitigation (AAM) algorithm. This algorithm, featuring early attack detection, reduces the AD processing workload during attacks, thereby decreasing the computational overhead on the Server, actively drops attack packets to reduce congestion, and in a timely manner and stops dropping packets after the attack ends to avoid the excessive loss of legitimate packets. 

The algorithm begins by testing the first window of $W > 0$ consecutive packets sequentially. The size of this window $W$, is experimentally chosen to optimize the accuracy of attack detection. If the AD identifies that a majority of the packets in the window are classified as "$ATTACK$," it concludes that an attack is occurring. It then drops the preceding $m + W$ packets to reduce congestion and skips $m > 0$ packets ahead in the incoming packet stream to test the next window of $W$ packets. This ensures that the AD system is not overwhelmed during an attack and can continue operating efficiently. Conversely, if the AD detects "$NO-ATTACK$", the current window of $W$ packets is forwarded to the Server, and the algorithm proceeds to test the next $W$ packets in the same manner, and the process is repeated. 

\subsection{Optimization of the Adaptive Attack Mitigation}\label{subsec:Optimization}

During a Flood Attack, a fraction ($0 < f \leq 1$) of incoming packets are part of the attack, while the remaining $1-f$ are benign, and $f$ is not known in advance. While dropping packets, the AAM will also drop some benign packets originating from various devices in the network. Although the AAM reduces the number of packets tested by the AD during an attack, it still introduces computational overhead by testing $W$ packets after every $m$-packet interval. Thus, in this subsection, we finalize the AAM by calculating the optimal $m$.

Let us denote by $X$ the total number of packets received at the SQF during an attack. Since $X$ is unknown in advance, it is treated as a random variable with the expected value $E[X]$. When AD initially identifies an attack within a $W$-packet window by identifying a majority of attack packets within the $W$-packet window, the first detection window serves as the starting point of the attack. The attack is considered to have ended when the AD detects a majority of non-attack packets in a subsequent $W$-packet window. The total number of detection windows $N$ during the attack, and its expected value $E[N]$, are therefore:
\begin{equation}
N=\lceil \frac{X-W}{m+W}\rceil,~E[N]\approx\frac{E[X]-W}{m+W}+\frac{1}{2},\label{approx}
\end{equation} 
where the expression for $E[N]$ is based on a mathematically proven \cite{ApproxFormula} first order approximation.

Since the SQF ensures that the AD processing time per packet remains constant at a value $\tau$, the Server overhead $\Omega$, and its expected value, are:
\begin{equation}
\Omega =N\tau W,~E[\Omega]\approx \tau W\big[\frac{E[X]-W}{m+W}+\frac{1}{2}\big]~.
\end{equation}
During the attack, the number of packets dropped by the AAM and its expected value, are:
\begin{equation}
\delta=W+N(m+W),~E[\delta]\approx E[X]+\frac{1}{2}(m+W).\nonumber
\end{equation}
Note that the dropped packets include those in the initial window classified as "ATTACK" and the following $m$ packets, in a pattern that repeats $N$ times. The final $W$-packet window is classified as "NO-ATTACK" and is not dropped. 

Within the $X$ packets which constitute the attack, we can assume that a fraction $0<f\leq 1$ are attack packets, but there may also be $X(1-f)$ non-attack packets. Since the last $\delta - X$ packets that are dropped after the attack ends contain only benign packets, 
the AS's overhead of reprocessing all the lost benign packets, denoted by $K$, assuming that they are all re-sent by their sources sequentially and tested by the AD, in windows of $W$ packets, will be:
\begin{eqnarray}
K= \tau W\lceil\frac{ fX+\delta -X}{W}\rceil,\nonumber\\
E[K]\approx\tau[ fE[X]+\frac{1}{2}m+W].
\label{dropped}
\end{eqnarray}
Thus the total average cost is:
\begin{equation}
C(AAM)=\alpha E[K]+\beta E[\Omega], \label{cost}
\end{equation}
where $\alpha>0$ is the importance we attribute to the cost of AD processing of packets that were lost during the attack and which came back after the attack ended,
while $beta>0$ is the importance we attribute to the AD processing of packets during the attack. Presumably, we should have $\beta>\alpha$ if, during an attack, the server is overloaded with other urgent tasks such as dropping of packets.
In addition, we know that during an attack, and despite the presence of the SQF, the actual processing of packets by the AD is increased.  Thus we would take $\frac{\beta}{\alpha}>1$ if we wish to reduce the overhead of AD processing during an attack, while we would take 
$\frac{\beta}{\alpha}<1$ if we wish to minimize the overhead throughout the system (including at the sources of traffic) caused by re-sending and re-processing the benign packets that were dropped by mistake during an attack.

Taking the derivative of the right-hand side of (\ref{cost}) with respect to $m$ we get:
\begin{equation}
\frac{1}{\tau}\frac{dC(AAM)}{dm}\approx \frac{1}{2}\alpha - \beta W
\frac{E[X]-W}{(m+W)^2}~,\label{der}
\end{equation}
and equating it to zero, we see that the total average cost $C(AAM)$ is approximately minimized when setting $m = m^*$:
\begin{equation}
	m^* \approx \sqrt{~2\frac{\beta}{\alpha}W[E[X]-W~]}~-W.\label{opt-m}
\end{equation}
We note that $m^* $ does not depend on $f$ and $\tau$, and it increases with the square root of $E[X]$.

Also, the minimum value of $C(AAM)$ computed at the value $m^*$ is:
\begin{eqnarray}
C^*(AAM)&\approx&\alpha\tau[fE[X]+\sqrt{~\frac{\beta}{2\alpha}W[~E[X]-W~]}+\frac{W}{2}]\nonumber\\
&&+\beta \tau W[\frac{E[X]-W}{\sqrt{~2\frac{\beta}{\alpha}W[~E[X]-W]}}+\frac{1}{2}]~.\label{costopt}
\end{eqnarray}

\begin{figure}[h!]
	\centering
	\includegraphics[height=7cm,width=9.5cm]{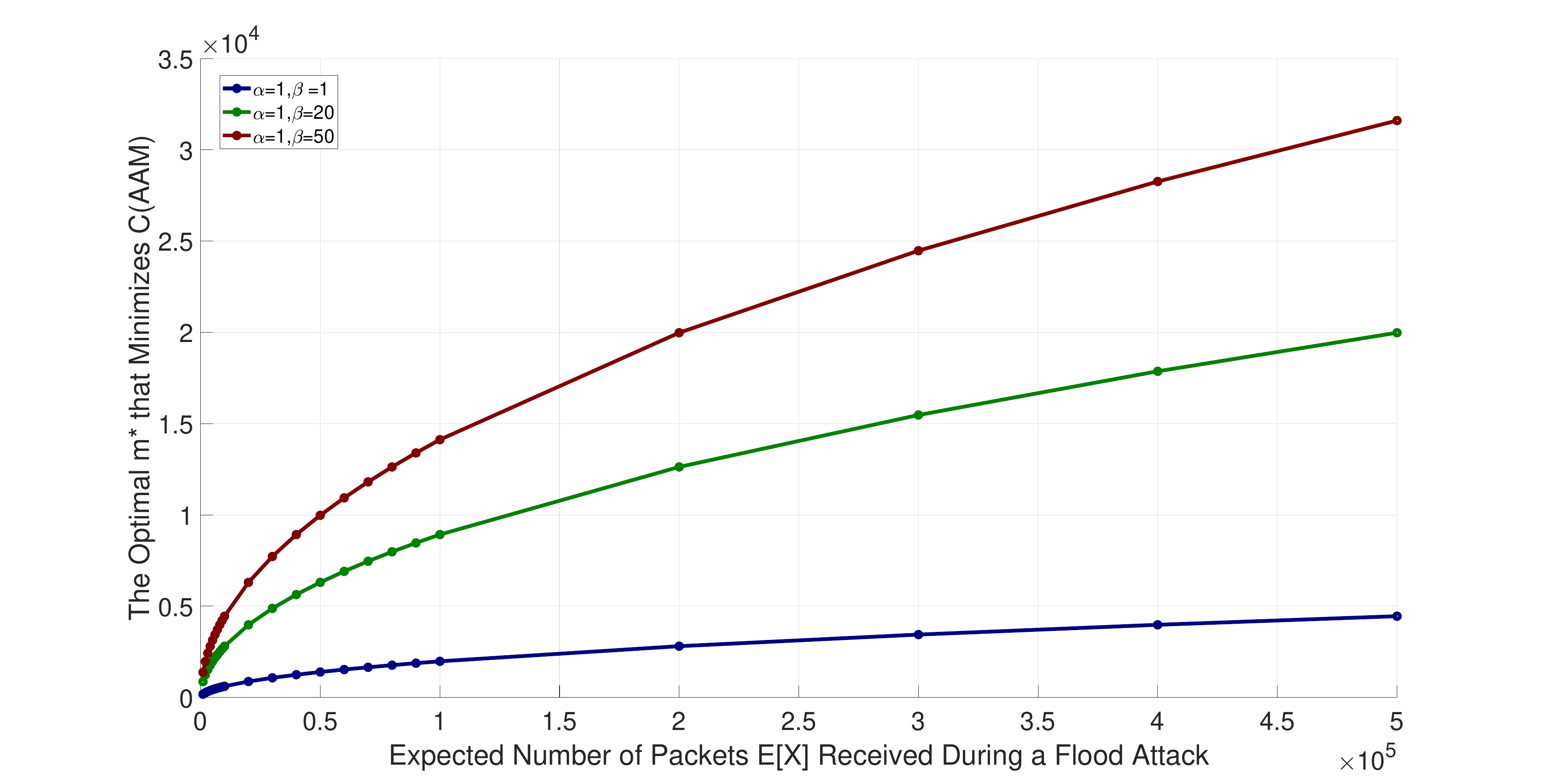}
	\caption{Theoretical graph of the optimum value $m^*$, which minimizes $C(AAM)$, as a function of $E[X]$ for $W=20$ and different values of $\frac{\beta}{\alpha}$.}
	\label{fig:optimization AAM}
\end{figure}

Figure \ref{fig:optimization AAM} shows $m^*$, the value that minimizes the cost $C(AAM)$,  for different values of $\frac{\beta}{\alpha}$ and $W=20$ (which is the actual value we have set for the AD in our experimental work), and different values of $E[X]$, 
Such figures can be used to choose the value of $m^*$ rapidly for different parameter sets.

\section{Results and Evaluation}\label{sec:Results}
In this section, we present the measurements and evaluation of the proposed system's behavior through real-time UDP Flood Attack experiments conducted on a test-bed. First, we describe the hardware and software configurations and settings of the test-bed prepared for the experiments. Then, we show the attack detection experiments conducted with and without SQF, followed by the experimental measurements of the optimization of the proposed AAM algorithm.

\subsection{Hardware  \& Software Configurations}\label{subsec:Hardware_Configurations}

\begin{figure}[t!]
	\centering
	\includegraphics[height=4cm,width=8.5cm]{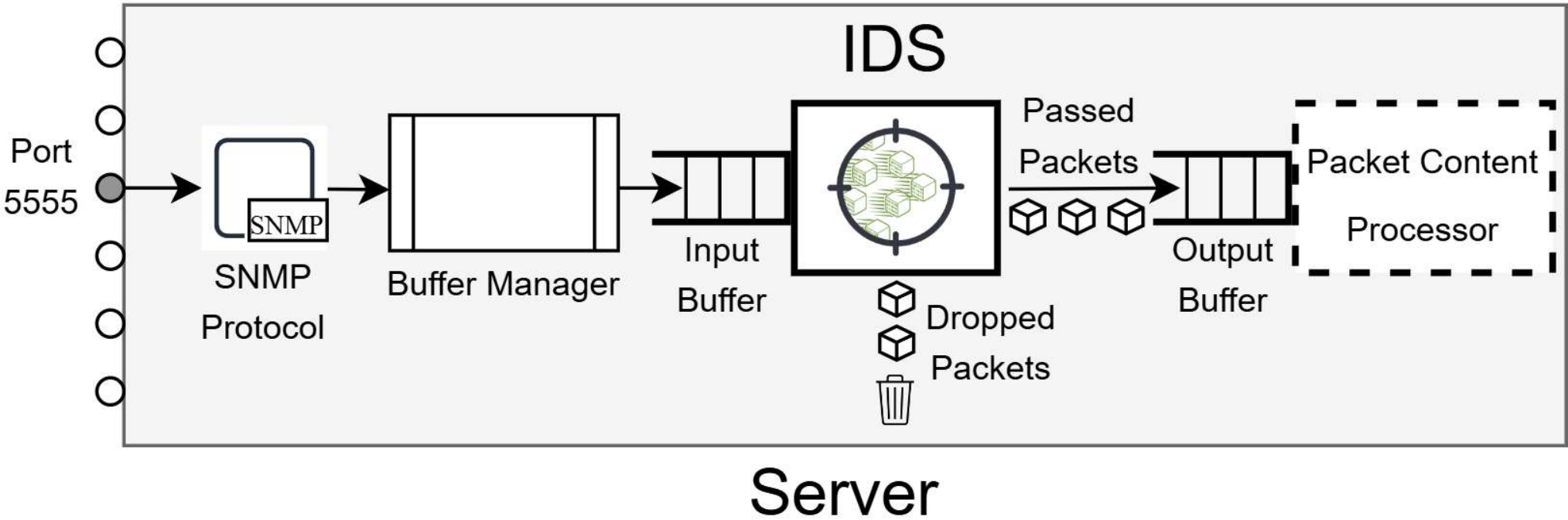}
	\caption{Schematic representation of the Server's software organization, featuring the simple network management processor (SNMP), the AD or Intrusion Detection System (labelled IDS in the figure), followed by the processing software for incoming data.}
	\label{Zero-1}
\end{figure}

The test bed on which we run experiments includes Raspberry Pi $4$ Model B Rev $1.2$ processors acting as mobile or sensor devices of the  pervasive system. Each one has a $1.5 GHz$ ARM Cortex-A$72$ quad-core processor, $2 GB$ RAM, and runs Raspbian GNU/Linux $11$. Some send attack traffic randomly or in a predetermined manner, while others send legitimate UDP packets with periodic machine temperature data to the Server. The devices have a network buffer of $176 KB$ and communicate with the Server via Ethernet as shown in Figure \ref{Zero-0}. The Gated Server is emulated by $3.1 GHz$ Intel $8$-Core $i7$-$8705$G processor, $16 GB$ RAM, and Linux 5.15.0 (Ubuntu SMP), receives packets at port $5555$ using the UDP protocol and processes them using SNMP $6.2.0$-$31$-generic as shown in Figure \ref{Zero-1}. Its NIC supports $1000 Mb/s$ speeds in full duplex mode, with a $208KB$ network buffer. We used UDP protocol due to its lightweight nature since it does not create connections and avoids ACKs \cite{UDP}.

The Maximum Transmission Unit (MTU) is set to $1.5 KB/packet$ for efficient packet transmission. Tests with $1000$ packets showed low latency, averaging $0.437$ ms, i.e., less than $15\%$ of the Server's AD processing time of $T_n \approx 3$ ms. So we neglect it in Section \ref{sec:Methodology}.
We used the MHDDoS public repository \cite{MHDDOS} for generating attack traffic that  includes $56$ real-world DoS emulators, enabling comprehensive testing with up-to-date scenarios.

\subsection{Attack Detection}\label{subsec:Attack_Detection}

\begin{figure}[h!]
	\centering
	\includegraphics[height=4cm,width=9cm]{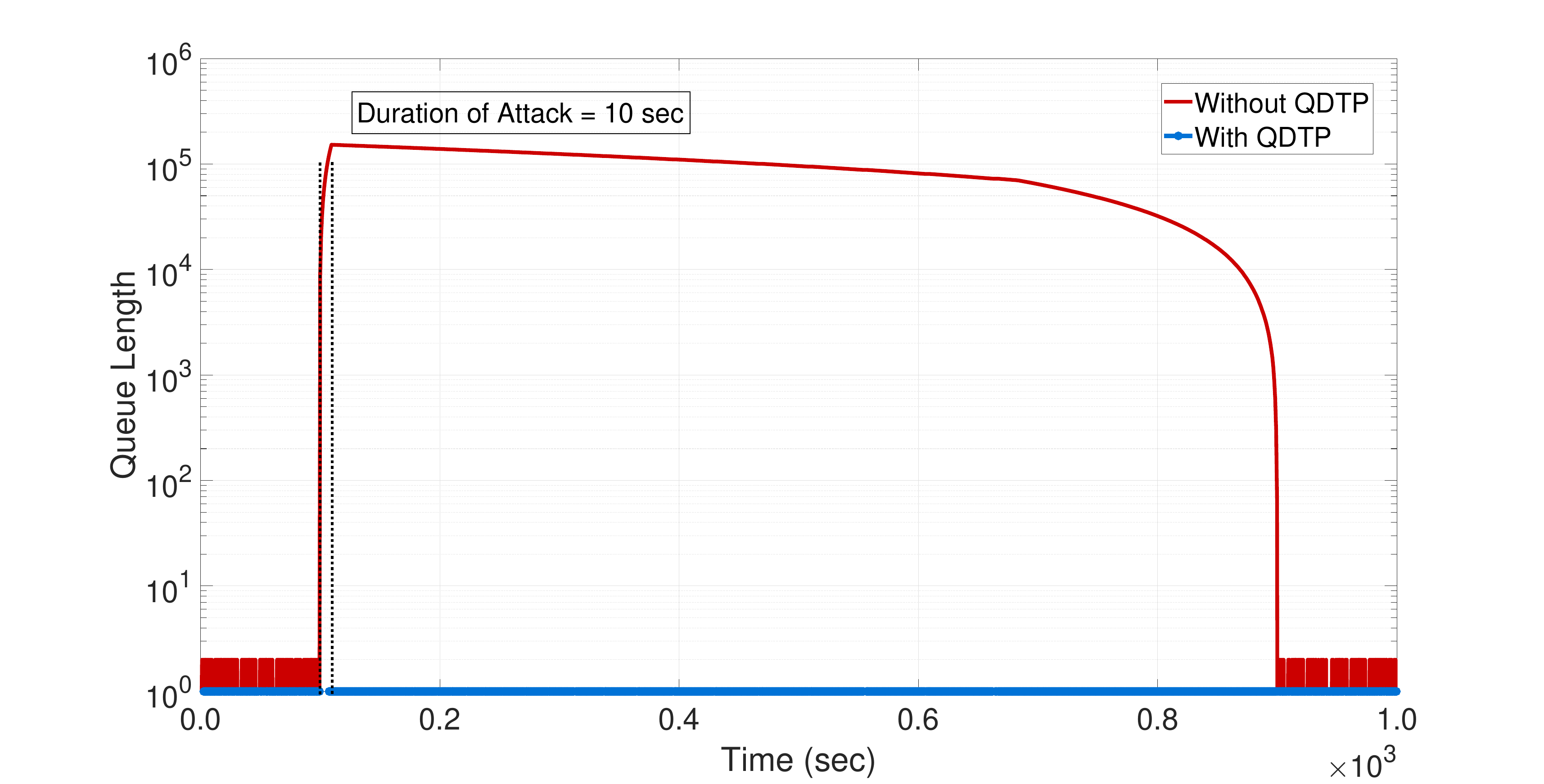}\\
	\includegraphics[height=4cm,width=9cm]{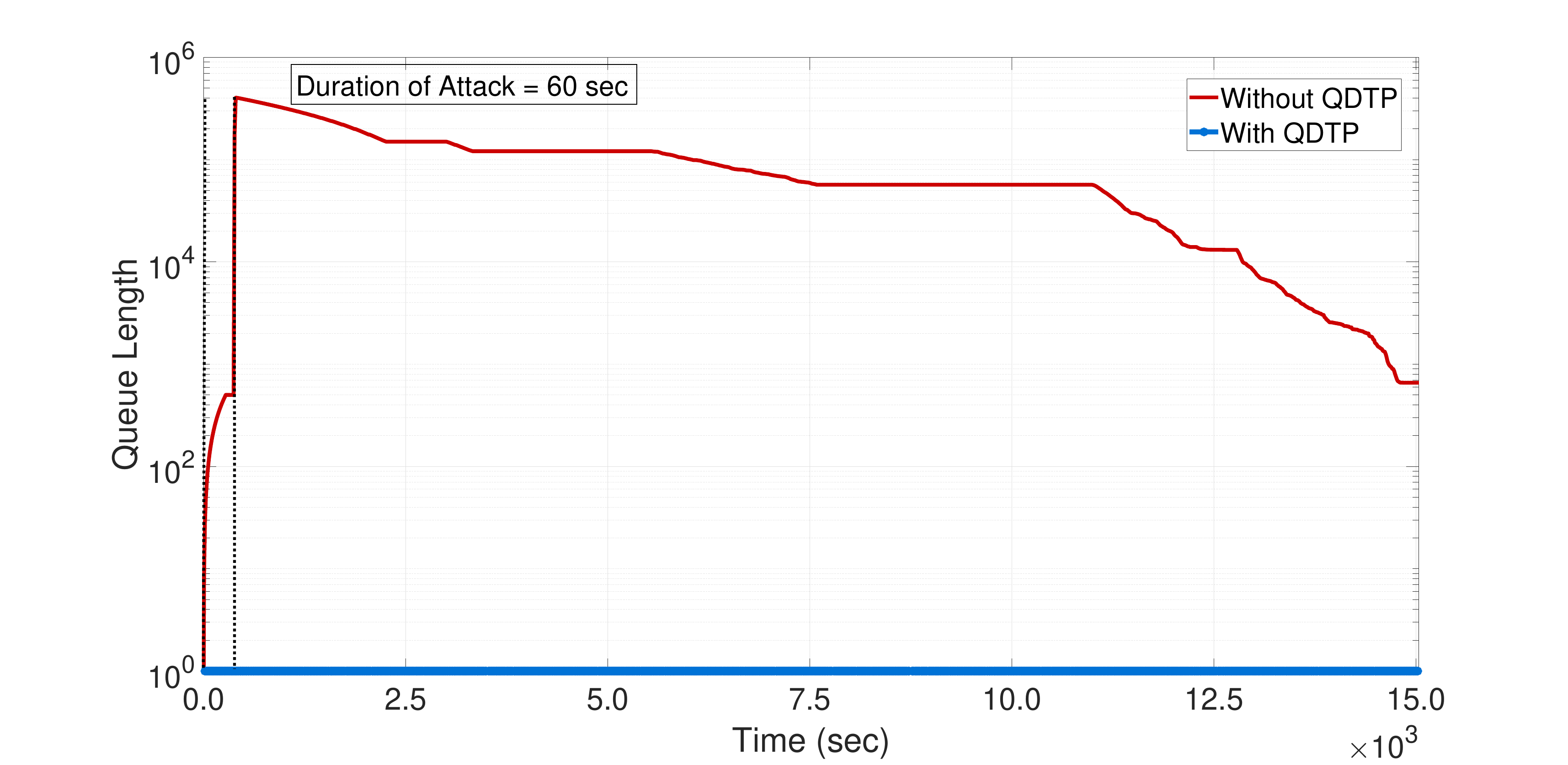}
	\caption{The Server queue length, measured experimentally and displayed on a logarithmic scale, is shown for a UDP Flood Attack lasting $10$ seconds (above) and $60$ seconds (below). The red curves represent the queue lengths without SQF, while the blue curves show the impact of SQF in reducing the queue length during both attacks duration, with $D = 3$ ms.}	
	\label{QL-10s&60s}
\end{figure}

Without the SQF, the Gateway Server is exposed to a huge accumulation of packets at the entrance of the AD during a Flood Attack, as shown by the experimental results shown in red in Figure \ref{QL-10s&60s} (above) and (below). Even if the attack lasts only a few seconds, as seen in the first experiment (above), a queue of approximately $153,667$ packets forms when the attack lasts $10$ seconds, and the Server requires around $15$ minutes to process them and resume normal operation. For  the $60$-second attack shown in the second experiment in the Figure below, a very large queue exceeding $400,000$ packets forms at the Server. This results in the system experiencing continuous interruptions during various periods, such as the period between minutes $133$ and $183$ of the experiment, where the Server becomes paralyzed and unable to operate the AD or other services. Furthermore, the system remains at constant risk of failure under these conditions.

In contrast, when SQF is used, the blue curves show that in both experiments, a short queue forms at the entrance of the Server, and packets are processed normally without delays or interruptions. This is because the parameter $D = 3$ ms is set very close to the average $T_n$ \cite{BestPaper}. Naturally, fluctuations in the Server's packet processing time result in the formation of a very small queue of no more than 20 to 30 packets.

\subsection{Attack Mitigation}\label{subsec:Attack_Mitigation}
This subsection presents the experimental evaluation of the proposed AAM algorithm that is described earlier in subsection \ref{subsec:Optimization}.

\begin{figure}[h!]
	\centering
	\includegraphics[height=6cm,width=9cm]{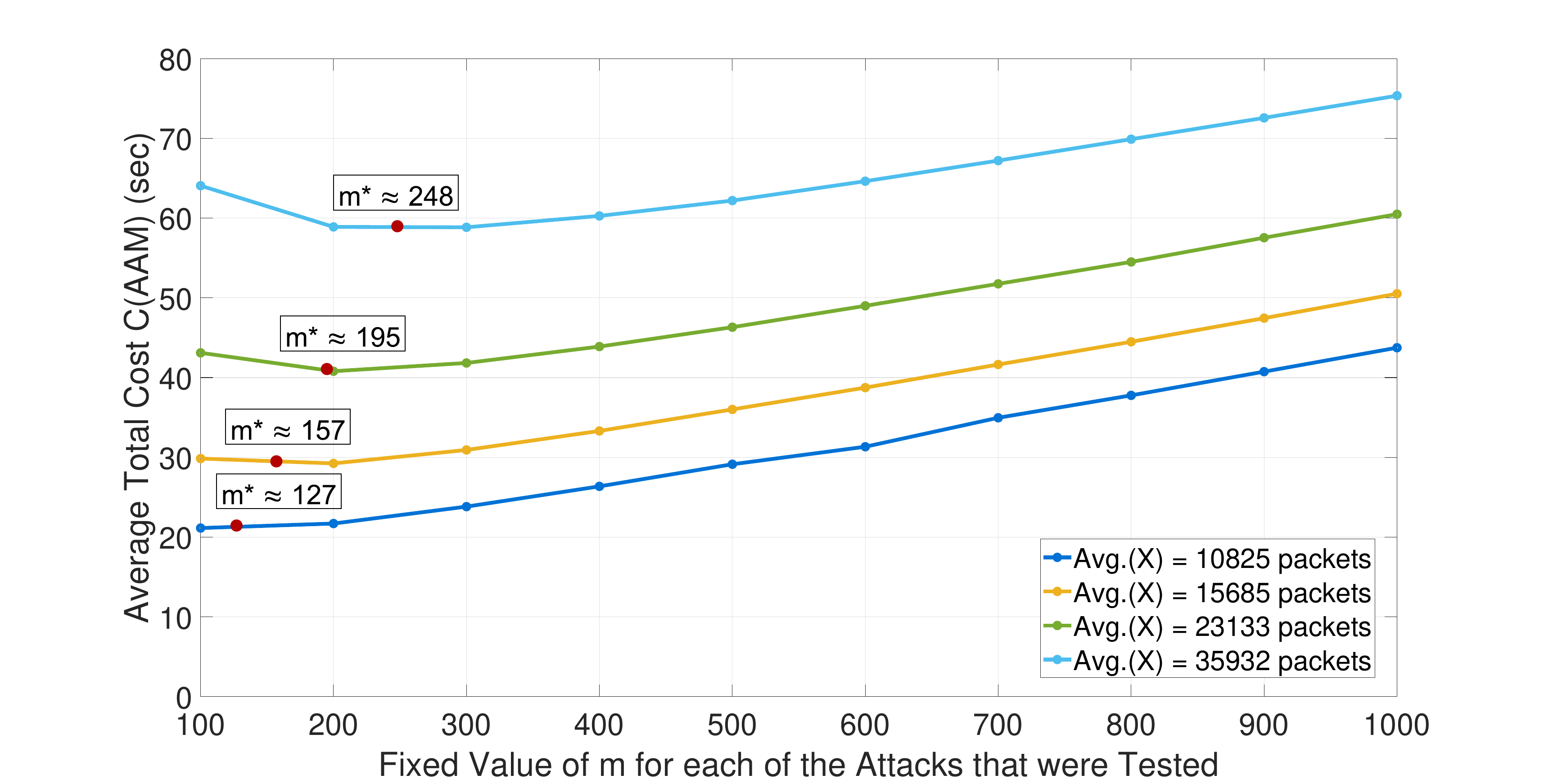}
	\caption{Experimental graph of the average total cost $C(AAM)$ against the parameter m.}
	\label{AverageOverhead}
\end{figure}

We conducted several experiments in which the parameter $X$ was randomly generated. To evaluate the performance of the AAM algorithm, each experiment was repeated $30$ times for a fixed expected number of packets $E[X]$ received during a Flood Attack. This process was repeated for different values of $E[X]$. The cost function $C(AAM)$ was calculated for each experiment, and its average value was computed by averaging the cost outcomes for different values of $X$ and each specific $E[X]$. The experimental measurements in Figure \ref{AverageOverhead} illustrate how the value of m increases with  $E[X]$, and empirically demonstrate the effectiveness of cost minimization by choosing $m^*$.

\begin{figure}[h!]
	\centering
	\includegraphics[height=5cm,width=9cm]{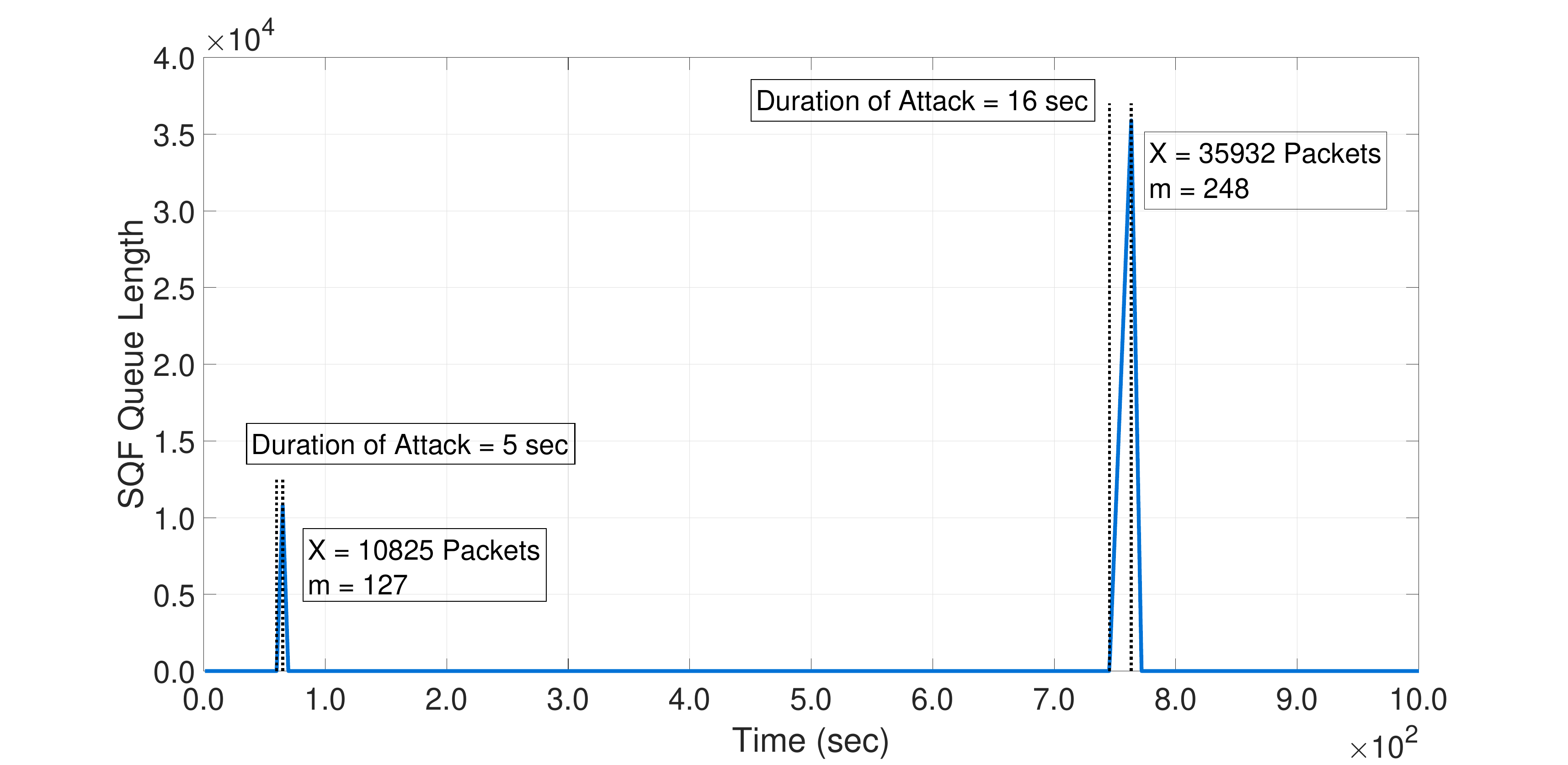}\\
	\includegraphics[height=5cm,width=9cm]{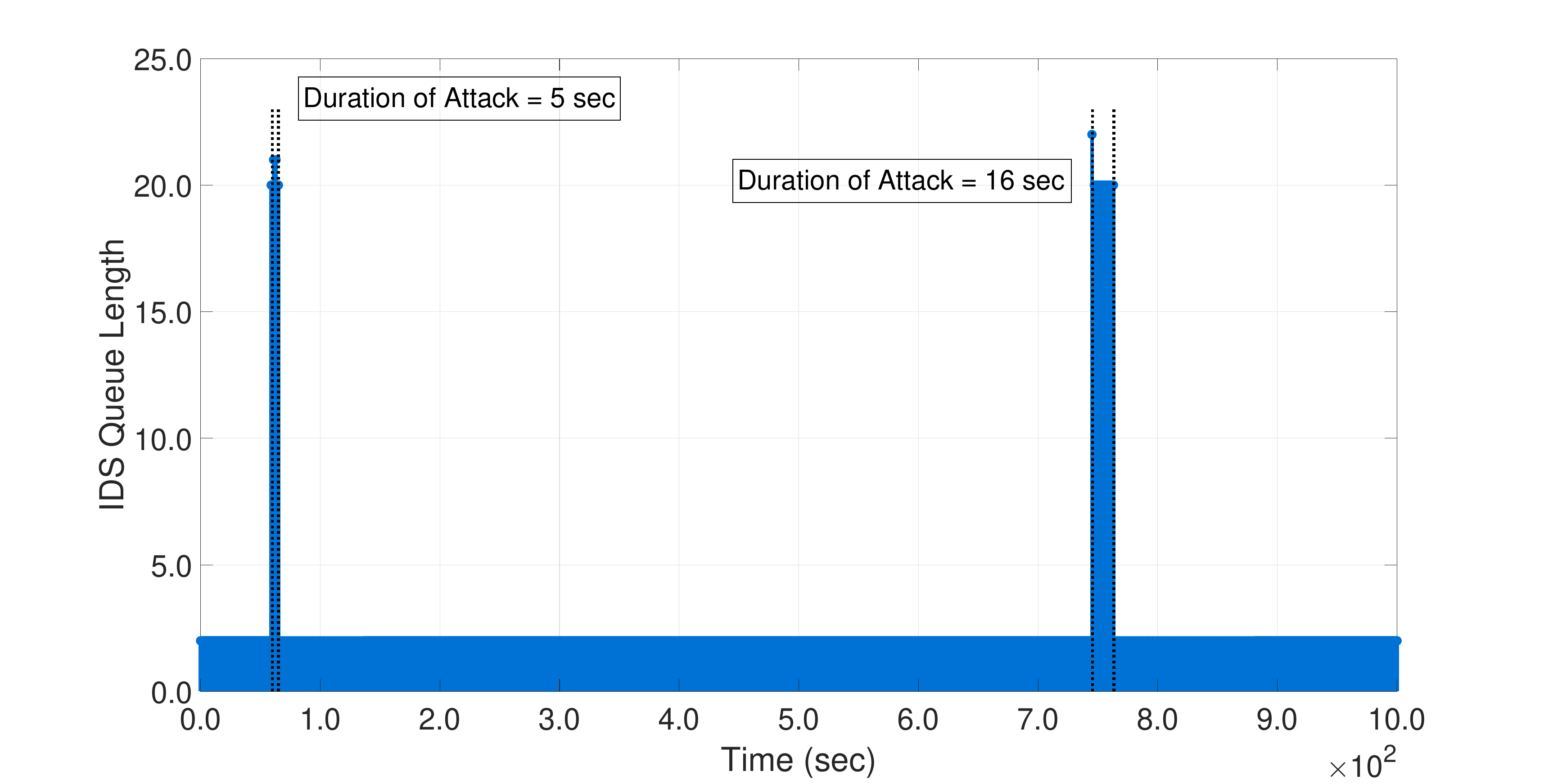}
	\caption{Timeline of the queue length measurements at the entrance of the SQF (above) for two successive attacks: the first involves approximately $10,000$ packets, and the second, about $40,000$ packets. At the entrance of the AD system, the combined use of SQF and AAM limits the input queue to around $20$ packets (below). When AAM activates, it computes $m^*$, which is $127$ for the first attack and $248$ for the second.}
	\label{AttackTimeLine_SQF_IDS}
\end{figure}

Figure \ref{AttackTimeLine_SQF_IDS} (above) shows the queue length at the entrance of the SQF during an experiment where the system is targeted by two Flood Attacks: the first containing approximately $10,000$ packets and the second $40,000$ packets. Upon receiving an alert from the AD indicating an attack, AAM calculates the value of $m^*$, which is $127$ for the first attack and $248$ for the second, as described previously by the theoretical formula. Figure \ref{AttackTimeLine_SQF_IDS} (below) shows the queue length at the entrance of the AD, which 
remains below $22$ packets, illustrating the effectiveness of the AAM in making rapid decisions to drop packets, even when two attacks occur consecutively within a short time frame.
Combining the SQF and the AAM ensures a low queue length at the AD immediately after an attack starts, and facilitates decision-making to drop packets and minimize $C(AAM)$.

	\section{Conclusions and Future Work}\label{sec:Conclusion}
	This paper sheds light on the impact of UDP Flood Attacks on resource-constrained Gateway Servers. It shows that even short-duration attacks can overwhelm the Server, causing prolonged backlogs and delays in normal operations. To address this issue, we discuss a new architecture including an SQF on a lightweight device implementing traffic shaping with the QDTP policy to protect the Server from congestion, and 
	an  AAM algorithm that adaptively samples and drops attack packets from the input stream, minimizing a cost function associated with benign packet drops and the sampling overhead. Experimental results show that SQF with AAM, effectively mitigate severe Flood Attacks.
	Future work will explore Pervasive Systems with multiple Gateways, and evaluate dynamic AD policies for complex Gateway networks with static and mobile nodes, and mitigation techniques that include traffic routing and dropping of attack packets. The energy consumption due to attacks will also be minimized.
\section*{Acknowledgement}
The support of the European Commission’s H2020 DOSS Project under GA No. 101120270 is gratefully acknowledged.	
	
	\bibliographystyle{IEEEtran}
	\bibliography{references,IDS,datasets,RNN,references_federated,references2,references3,references4,mybib,security_issues_references,references_arnn_conference,IoVSecurity_references}

\end{document}